\documentclass[proof]{pasj01}
\SetRunningHead{Shigeyama et al.}{Repulsion of fall back matter}
\usepackage{times}

\begin{document}

\title{Repulsion of fallback matter due to central energy source in supernova}
\author{Toshikazu Shigeyama\altaffilmark{1}, Kazumi Kashiyama\altaffilmark{1,2}}
\altaffiltext{1}{Research Center for the Early Universe, Graduate School of Science, University of Tokyo, Bunkyo-ku, Tokyo 113-0033}
\altaffiltext{2}{Department of Physics, Graduate School of Science, University of Tokyo, Bunkyo-ku, Tokyo 113-0033, Japan}


\maketitle

\begin{abstract}
The flow of fallback matter being shocked and repelled back by an energy deposition from a central object is discussed by using newly found self-similar solutions. 
We show that there exists a maximum mass accretion rate if the adiabatic index of the flow is less than or equal to 4/3. 
Otherwise we can find a solution with an arbitrarily large accretion rate by appropriately shrinking the energy deposition region. 
Applying the self-similar solution to supernova fallback, we discuss how the fate of newborn pulsars or magnetars depends on the fallback accretion and their spin-down power. 
Combining the condition for the fallback accretion to  bury the surface magnetic field into the crust, 
we argue that supernova fallback with a rate of $\dot M_{\rm fb} \sim 10^{-(4\mbox{-}6)}\,M_\odot\,\rm s^{-1}$ could be the main origin of the diversity of Galactic young neutron stars,
i.e., rotation-powered pulsars, magnetars, and central compact objects. 
\end{abstract}

\section{Introduction}
Compact objects such as neutron stars and black holes are formed in collapsing stars some of which eventually end up with supernova explosions. 
In general, a fraction of supernova ejecta falls back toward a newborn compact object \citep{1971ApJ...163..221C, 1972SvA....16..209Z, 1988Natur.333..644M}. 
The accretion rate and its temporal evolution can be determined by e.g., the strength of the supernova shock and the progenitor structure (e.g., \cite{2016ApJ...818..124E}). 
On the other hand, newborn compact objects can continuously deposit energy into the fallback matter via e.g., neutrino emission, pulsar activity and/or accretion-disk wind (e.g., \cite{2011ApJ...736..108P}). 
Competition between the fallback accretion onto and the energy deposition from compact objects may result in a variety of outcomes and lead to a diversity of compact objects. 

In particular, the dynamics of the fallback accretion may be related to the diversity of relatively young ($\sim 1\mbox{-}10~\rm kyr$) neutron stars; 
there are three distinct populations, ordinary pulsars, magnetars, and central compact objects (CCOs)~\footnote{They sometimes show links to another population.  
For example, some pulsars underwent soft-gamma repeater like bursts \citep{2010ApJ...711L...1V}, 
and a CCO was found to exhibit magnetar-like activity \citep{2016MNRAS.463.2394D, 2016ApJ...828L..13R}.}. 
Based on the spectral and timing observations, they are considered to have different energy sources, 
the rotation, magnetic-field, and thermal energies, respectively.
The most distinctive difference between them is the magnetic-field strength; young pulsars typically have dipole fields of $B_{\rm d} \lesssim 10^{13}$ G~\footnote{http://www.atnf.csiro.au/research/pulsar/psrcat/}.
Magnetars are considered to be powered by decays of stronger fields of $B_* >\mathrm{a~few}\times10^{13}$ G~\citep{2014ApJS..212....6O} 
while CCOs have considerably weaker dipole fields of $B_{\rm d} \ll 10^{12}$ G~\citep[and so on]{2001ApJ...559L.131P, 2006ApJ...653L..37P}.
Although the diversity of the magnetic field strength can be attributed to conditions before and during the neutron star formation,
i.e., the magnetic field strength of the iron core of a progenitor (the fossil scenario; \cite{2006MNRAS.367.1323F}) or the magnetic field amplification by a dynamo process~\citep{1992ApJ...392L...9D,2015Natur.528..376M}, 
it may be also possible to attribute the diversity to conditions after the formation, i.e., various rates of the fallback accretion onto the neutron star. 

If the accretion rate is sufficiently high, the fallback matter can compress and bury the magnetic field into the neutron star crust~(e.g., \cite{2010RMxAA..46..309B,2016MNRAS.456.3813T}).
Such a case may correspond to a CCO formation (the hidden magnetic field scenario; \cite{1995ApJ...440L..77M,1995ApJ...442L..53Y}). 
For example, \citet{2016MNRAS.456.3813T} obtained the critical accretion rate for burying the magnetic field 
by considering the competition between the magnetic pressure of the surface field and the ram pressure of the fallback matter. 
In these studies, however, the energy deposition from the neutron star has been neglected. 
If the energy deposition rate is sufficiently large, the fallback matter can be repelled before reaching the surface.
The critical conditions can be obtained by considering the competition between the fallback accretion onto and the energy deposition from the neutron star.
This may define additional bifurcations of the neutron star population.

To this end, we here investigate the dynamics of fallback matter being pushed back by an energy deposition from the central object with newly constructed spherically symmetric self-similar solutions. 
We consider fallback matter marginally bound by the gravitational field of the central object with a mass $M_{\rm c}$ 
and a power-law energy deposition rate from the central object;  
\begin{equation}\label{eq:edep}
\dot{Q}=L_l t^l.
\end{equation}
Here $L_l$ and $l$ are constants and $t$ is the time measured from the onset of the energy deposition. 
Although our solutions are described by a single dimensionless variable 
\begin{equation}
\xi=\frac{r}{(GM_{\rm c}t^2)^{1/3}},
\end{equation}
where $r$ is the radial coordinate and $G$ denotes the gravitational constant, 
a variety of density and velocity structures can be realized depending on the strength of the deposition and so on, 
which, to our knowledge, have not been seen in previous studies. 

The structure of the paper is as follows. 
The next  section describes our model. 
Section 3 presents our self-similar solutions with various values of parameters. 
In Section 4, we discuss some applications of our solutions to neutron star formation. Section 5 summarizes the results and discusses relations to previous works.  

\begin{figure}
\begin{center}
  \includegraphics[width=0.7\textwidth]{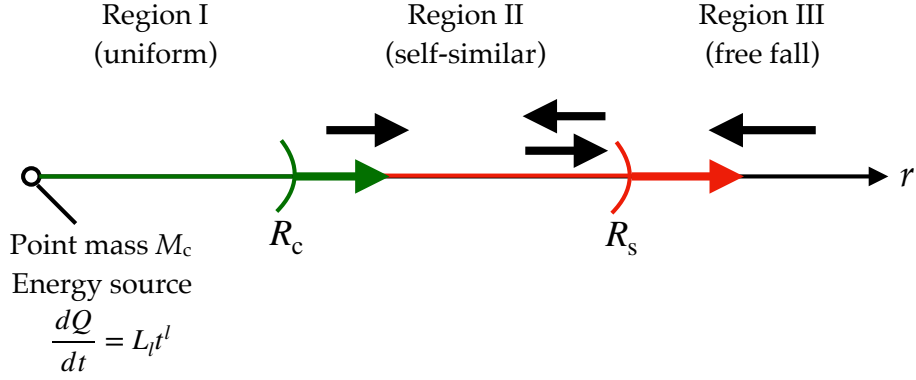}
\end{center}
\label{fig:regions}
\caption{Schematic view of the model consisting of three regions. Region I is adjacent to the central object. The energy from the central object is supposed to be uniformly deposited in this region. Region II contains the shocked fallback matter and is separated from Region I by a contact surface. In Region III, the matter freely falls due to the gravity of the central object. An expanding shock front separates Region II from Region III. The black arrows indicate the direction of the flows while the green and red arrows indicate the propagations of the contact surface and the shock front, respectively. Note that the direction of the flow immediately behind the shock front depends on model parameters. See section \ref{sec:results} for details.}
\end{figure}

\section{Model}\label{sec:model}
We consider the flow of fallback matter affected by the energy deposition from a central object. 
Following \citet{2016PASJ...68...22M}, we divide the flow into three distinct regions (see Fig. \ref{fig:regions}): 
the innermost region  (Region I) adjacent to the central object where the energy is deposited uniformly,  Region II where the fallback matter is shocked, and the outermost region in which cold matter freely falls due to the gravity of the central object (Region III). Regions I and II are separated by the contact discontinuity specified by the dimensionless variable introduced above as $\xi=\xi_{\rm c}$. The fluid in Region II has a constant adiabatic index $\gamma$ and turns into outflow at a certain point depending on $\xi_{\rm c}$ and $\gamma$.   
The shock front divides the fallback matter into Regions II and III at $\xi=\xi_{\rm s}$. 
Such a flow structure across Regions I and II has been reproduced by numerical simulations~(e.g., \cite{2016PASJ...68...22M}).

\subsection{Energy deposition region (Region I)}
The central object is assumed to  deposit energy in a uniform spherical region with a radius $R_{\rm c}=\xi_{\rm c}(GM_{\rm c}t^2)^{1/3}$. The internal energy $E_{\rm c}$ and pressure $P_{\rm c}$ in this region evolve with time $t$ according to the first law of thermodynamics as
\begin{equation}\label{eq:eir}
\frac{dE_{\rm c}}{dt}+P_{\rm c}\frac{d}{dt}\left(\frac{4\pi R_{\rm c}^3}{3}\right)=L_l t^l.
\end{equation}
 Here $L_l$ and $l$ are constant. The subscript $l$ of $L_l$ indicates that the dimension of this quantity depends on the value of $l$. This region becomes very hot and the equation of state can be approximated by that for ultra-relativistic gas, that is, $E_{\rm c}=4\pi P_{\rm c} R_{\rm c}^3$.

\subsection{Accreted matter (Regions II and III)}
Accreted matter can be divided into the shocked fallback matter (Region II) and the surrounding accreted matter (Region III). The motion of the surrounding matter is not affected by the energy deposition but controlled solely by the gravity from the central object.  We describe the flow in Region II  by using  a self-similar solution.
We note that the problems treated in this paper involve two constant quantities with physical dimensions $GM_{\rm c}$ and $L_l$. 

Here we assume that the spherically symmetric flow depends on time $t$ and $r$ only through $t$ and $\xi(r,\,t)$. Thus the density $\rho$, the velocity $v$, and the pressure $p$ take the forms of
\begin{eqnarray}
&\rho({\xi},{t})=\frac{L_l t^{l-\frac{1}{3}}{D}(\xi)}{\left(GM_{\rm c}\right)^{5/3}},&\label{eq:density} \\ 
&v({\xi},{t})=\left(\frac{{GM_{\rm c}}}{t}\right) ^{1/3}V(\xi) ,& \label{eq:velocity} \\
& p({\xi},{t})=\frac{L_l  t^{l-1}\Pr (\xi)}{GM_{\rm c} },& \label{eq:pressure}
\end{eqnarray}
as functions of $\xi$ and $t$. Here we have introduced dimensionless functions $D(\xi)$, $V(\xi)$, and $\Pr(\xi)$. The governing equations are described as
\begin{eqnarray}
&\frac{\partial\rho(\xi, t)}{\partial t}+\frac{\partial r^2\rho(\xi, t) v(\xi, t)}{{r^2}\partial r}=0,& \\
&\rho(\xi, t) \left(\frac{\partial v(\xi, t)}{\partial t}+v(\xi, t) \frac{\partial v(\xi, t)}{\partial r}\right)+\frac{GM_{\rm c}\rho(\xi, t)}{r^2}+\frac{\partial p(\xi, t)}{\partial r}=0, &\\
&\left(\frac{\partial }{\partial t}+v(\xi, t) \frac{\partial }{\partial r}\right)\left[\frac{p(\xi, t)}{(\gamma-1)\rho(\xi, t)}\right]+p(\xi, t) \left(\frac{\partial }{\partial t}+v(\xi, t) \frac{\partial }{\partial r}\right)\left[\frac{1}{\rho(\xi, t)}\right]=0. &
\end{eqnarray}
After some manipulations we obtain ordinary differential equations for dimensionless functions $D(\xi)$, $V(\xi)$, and $\Pr(\xi)$ as
\begin{eqnarray}
&\xi (2 \xi-3 V(\xi)) D^\prime(\xi)-D(\xi) \left(3 l \xi+3 \xi V'(\xi)+6 V(\xi)-\xi\right)=0,&\label{eq:ssflc}\\
&D(\xi) \left(\xi^2 (2 \xi-3 V(\xi)) V^\prime(\xi)+\xi^2 V(\xi)-3\right)=3 \xi^2 \Pr ^\prime(\xi),&\label{eq:ssflm}\\
&{\gamma(2 \xi-3 V(\xi)) \frac{D^\prime(\xi)}{D(\xi)}+(3 (1-\gamma) l+\gamma-3)+(3 V(\xi)-2 \xi) \frac{\Pr ^\prime(\xi)}{ \Pr (\xi)}}=0,\label{eq:ssfle}&
\end{eqnarray}
where $^\prime$ denotes the derivative with respect to $\xi$. 
Eqs. (\ref{eq:ssflc}-\ref{eq:ssfle}) can be rewritten as 
\begin{eqnarray}
&D'(\xi)= \frac{D(\xi) \left(D(\xi)  \left(9-3 \xi V(\xi) ((3 l-4) \xi+6 V(\xi))+(6 l-2) \xi^3\right)\right)}{\xi^2 \left(D(\xi) (2 \xi-3 V(\xi))^2-9 \gamma \Pr (\xi)\right)}&\nonumber \\
&+\frac{ 9 \xi^2 (-3 \gamma l+\gamma+3 l-3) D(\xi)\Pr (\xi)}{\xi^2 (2 \xi-3 V(\xi)) \left(D(\xi) (2 \xi-3 V(\xi))^2-9 \gamma \Pr (\xi)\right)},& \label{eq:ssd}\\
&V'(\xi)= \frac{9 \xi \Pr (\xi) (2 \gamma V(\xi)+(l-1) \xi)-D(\xi) (2 \xi-3 V(\xi)) \left(\xi^2 V(\xi)-3\right)}{\xi^2 \left(D(\xi) (2 \xi-3 V(\xi))^2-9 \gamma \Pr (\xi)\right)},&  \label{eq:ssv}\\
&\Pr '(\xi)= \frac{3 D(\xi) \Pr (\xi) \left(3 \xi V(\xi) (\xi (\gamma-l+1)-2 \gamma V(\xi))+3 \gamma+2 (l-1) \xi^3\right)}{\xi^2 \left(D(\xi) (2 \xi-3 V(\xi))^2-9 \gamma \Pr (\xi)\right)}.  \label{eq:ssp}&
\end{eqnarray}
From the second term of the right hand side of equation (\ref{eq:ssd}), we can find that the derivative of the density diverges at  $\xi_{\rm c}$ where
\begin{equation}\label{eq:contact}
V(\xi_{\rm c})=\frac{2\xi_{\rm c}}{3},
\end{equation}
is satisfied while the other derivatives of the pressure and the velocity do not diverge. Thus this point $\xi=\xi_{\rm c}$ defines the contact surface between Regions I and II.
To obtain a solution, we numerically integrate Eqs.(\ref{eq:ssd}-\ref{eq:ssp}) from $\xi=\xi_{\rm s}$ to $\xi=\xi_{\rm c}$.

The solution for the flow in Region III ($\xi\geq\xi_{\rm s}$) is obtained by ignoring pressure in equations (\ref{eq:ssflc}) and (\ref{eq:ssflm}) (or (\ref{eq:ssd}) and (\ref{eq:ssv})) as
\begin{eqnarray}
&V(\xi)=-\sqrt{\frac{2}{\xi}},& \label{eq:fbv}\\
&D(\xi)=D_{\rm fb}\exp{\left[\int_{\xi_{\rm s}}^\xi dx\frac{2(3l-1)x-9\sqrt{2}x^{-1/2}}{2x(2x+3\sqrt{2}x^{-1/2})}\right]}.& \label{eq:fbd}
\end{eqnarray}
Here a constant $D_{\rm fb}$  denotes $D(\xi_{\rm s})$ in the un-shocked flow (Region III). 
It follows from substitutions of these expressions into equations (\ref{eq:density}) and (\ref{eq:velocity}) that the distributions of the density and velocity in the fallback matter evolve as
\begin{eqnarray}
&\rho(r,\,t)=\frac{L_lD_{\rm fb}t^{l-1/3}}{(GM_{\rm c})^\frac{5}{3}}\exp{\left[\int_{\xi_{\rm s}}^\frac{r}{(GM_{\rm c}t^2)^{1/3}} dx\frac{2(3l-1)x-9\sqrt{2}x^{-1/2}}{2x(2x+3\sqrt{2}x^{-1/2})}\right]},& \\
&v(r,\,t)=-\sqrt{\frac{2GM_{\rm c}}{r}} \label{eq:fbvel}, &
\end{eqnarray}
and that the density approaches a power-law function proportional to $r^{(3l-1)/2}$ independent of $t$ in the limit of $r\rightarrow\infty$. Thus the flow approaches a stationary state for a large $r$ ($\xi$).  
The velocity expressed by equation (\ref{eq:fbvel}) implies that the fallback matter is initially at rest at  large distances from the center. {Thus the total energy of the flow in Region III is equal to zero while the energy of the flow in Region II becomes positive due to the energy supply from the central source. This  indicates that all the solutions presented here describe the flow eventually repelled by the energy supply.} Because the density distribution depends on the parameter $l$ that specifies the nature of the central energy source, this density distribution is required to lead to a self-similar solution in which the shock front expanding with the radius proportional to $t^{2/3}$ for a given energy source. If the central source is more energetic, which corresponds to larger $L_l$, the fall back matter can be denser, while if the gravity of the central object is stronger, which corresponds to a larger $M_{\rm c}$, then the fallback matter needs to be more sparse to have an expanding shock front.

We ignore physical processes like neutrino heating and cooling, radiative diffusion, photo-disintegration of nuclei, and electron-positron pair production and annihilation, some of which might play crucial roles in the fate of the fallback matter. Instead, we have presented solutions with varying the adiabatic index and the exponent $l$ of $\dot Q$ in equation (\ref{eq:edep}), which mimic some of the effects of these processes.

\subsection{Boundary conditions}
\subsubsection{Contact surface}\label{sec:cs}
We require a condition that the flow has a continuous pressure distribution through the contact surface between Regions I and II where equation (\ref{eq:contact}) holds. From the evolution of the pressure derived from equation (\ref{eq:eir}), this requirement leads to 
\begin{equation}\label{eq:pc}
\Pr(\xi_{\rm c})= \frac{3}{4 \pi (5 +3 l) \xi_{\rm c}^3}.
\end{equation}
This boundary condition is equivalent to the condition that the energy deposited by the central source is equal to the total energy of the flow inside the shock front. 
Note here that the accreted matter entering into the shock front carries no energy (the sum of the energy densities of the matter is $v^2(r,\,t)/2-GM_{\rm c}/r=0$ from equation \ref{eq:fbvel}). 
We found that the self-similar flow around the contact surface behaves as
\begin{eqnarray}
&D(\xi)\sim D_{\rm c}(\xi-\xi_{\rm c})^\frac{3 - \gamma -3 l + 3 \gamma l}{3 ( 2 \gamma -1+ l)},& \label{eq:csd}\\
&V(\xi)\sim\frac{2\xi_{\rm c}}{3} + \frac{3 - 4\gamma - 3 l}{3\gamma} (\xi - \xi_{\rm c}), &
\end{eqnarray}
where $D_{\rm c}$ is a constant. From equation (\ref{eq:csd}), the density distribution near the contact surface drastically changes if $\gamma>3/2$ or not. If $\gamma<3/2$, the density decreases to zero toward the contact surface. If $\gamma>3/2$, solutions with $l=(-3 + \gamma)/[3 ( \gamma-1)]$ give constant densities at the contact surface between Regions I and II. Otherwise the density at the contact surface diverges to infinity ( $l<(-3 + \gamma)/[3 ( \gamma-1)]$) or goes to 0  ( $l>(-3 + \gamma)/[3 ( \gamma-1)]$).

\subsubsection{Shock front}\label{sec:front}
We assume that a strong shock propagates in the accreted matter. The location of the shock front is specified by $\xi=\xi_{\rm s}$, where $\xi_{\rm s}$ is a constant. The Rankine-Hugoniot jump conditions at the shock front can be written as
\begin{eqnarray}
&D(\xi_{\rm s})=\frac{\gamma+1}{\gamma-1}D_{\rm fb},\label{eq:shockd}&\\
&V(\xi_{\rm s})=\frac{1-\gamma}{\gamma+1}\sqrt{\frac{2}{\xi_{\rm s}}}+\frac{4\xi_{\rm s}}{3\left(\gamma+1\right)},&\label{eq:shockv}\\
&\Pr(\xi_{\rm s})=\frac{4D_{\rm fb}\sqrt{\xi_{\rm s}^{3(l-1)}}}{\gamma+1}\left(\frac{\sqrt{2\xi_{\rm s}^3}}{3}+1\right)^2.&
\end{eqnarray}
Here we have used the fact that the dimensionless velocity of the accreted matter in Region III at the shock front is  $-\sqrt{2/\xi_{\rm s}}$ (see eq. \ref{eq:fbv}) and that the dimensionless density at the shock front is denoted by  $D_{\rm fb}$. 
The value of  $D_{\rm fb}$ is determined to satisfy the boundary condition (\ref{eq:pc}) at the contact surface for each given $\xi_{\rm s}$.  
\begin{figure}
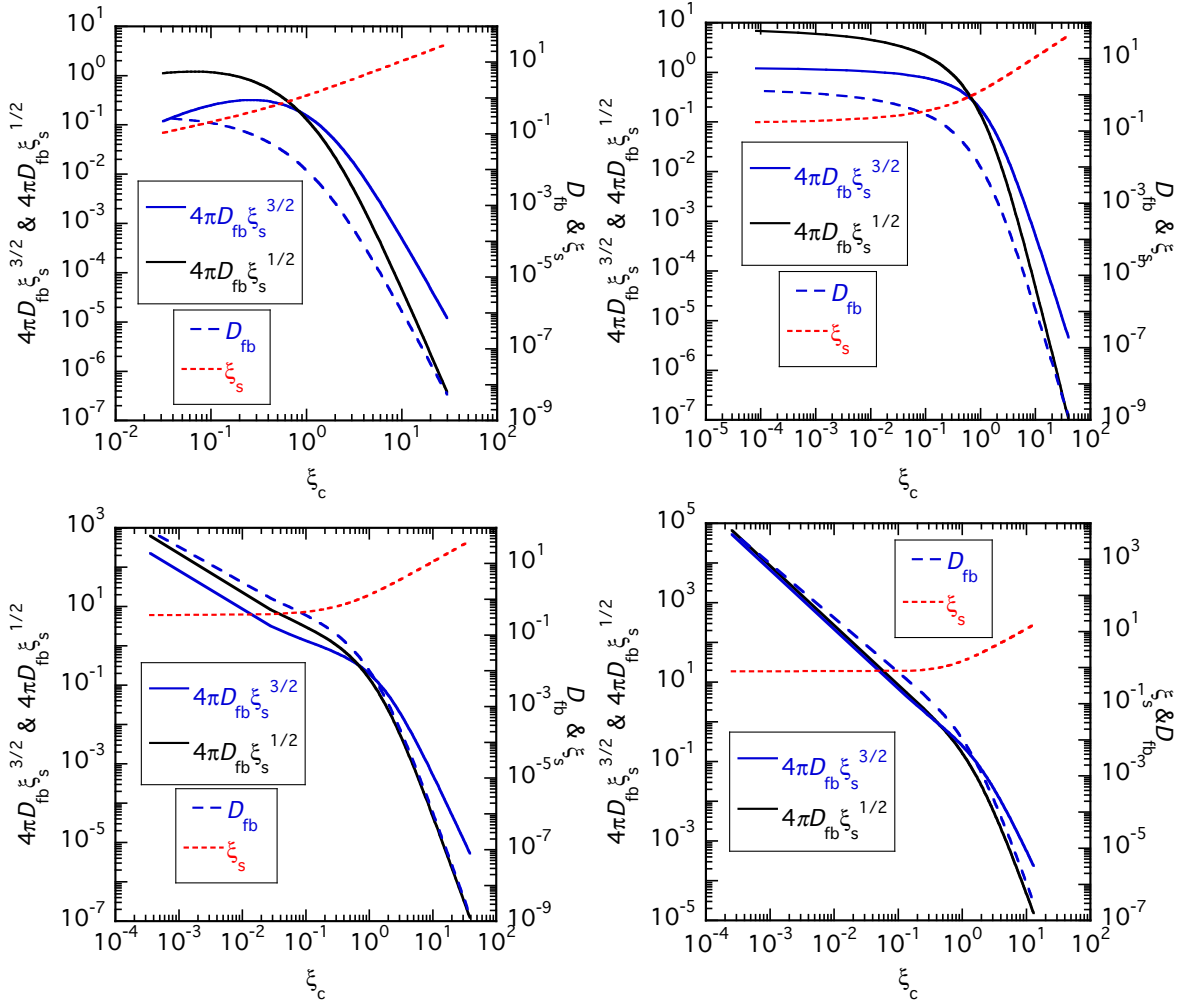

\begin{center}
 \includegraphics[width=0.45\textwidth]{g65l0.eps} \includegraphics[width=0.45\textwidth]{g43l0.eps}
 \includegraphics[width=0.45\textwidth]{g75l0.eps}  \includegraphics[width=0.45\textwidth]{g53l0.eps}
\caption{Relations between some characteristic quantities and $\xi_{\rm c}$ for solutions with four different values of $\gamma$: $6/5$ (top left panel), $4/3$ (top right panel), $7/5$ (bottom left panel),  and $5/3$ (bottom right panel). $l=0$ for all the solutions.}
\label{fig:accretion}
\end{center}
\end{figure}

\begin{figure}
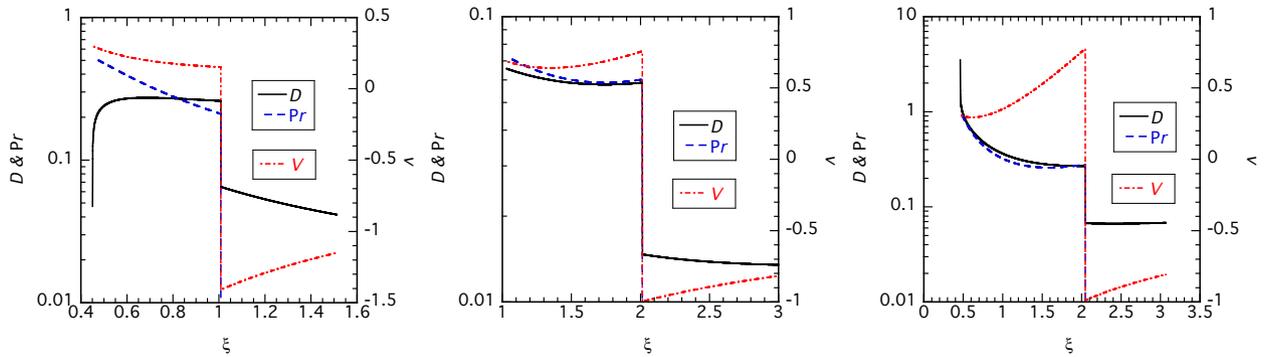

\begin{center}
  \includegraphics[width=0.32\textwidth]{dist_g53_l0_xs1.eps} \includegraphics[width=0.32\textwidth]{dist_g53_l-23_xs2.eps} \includegraphics[width=0.32\textwidth]{dist_g53_l-0.9_xs2.eps}
\caption{Distributions of the density $D$, the pressure $\Pr$, and the velocity $V$ as functions of $\xi$ for solutions with 3 different values of $l$: 0  (left panel), $-2/3$ (middle panel), and $-0.9$ (right panel). $\gamma=5/3$ for all the solutions. The values of $\xi_{\rm s}$ are chosen arbitrarily.}
\label{fig:dist53}
\end{center}
\end{figure}

\begin{figure}
\begin{center}
  \includegraphics[width=0.45\textwidth]{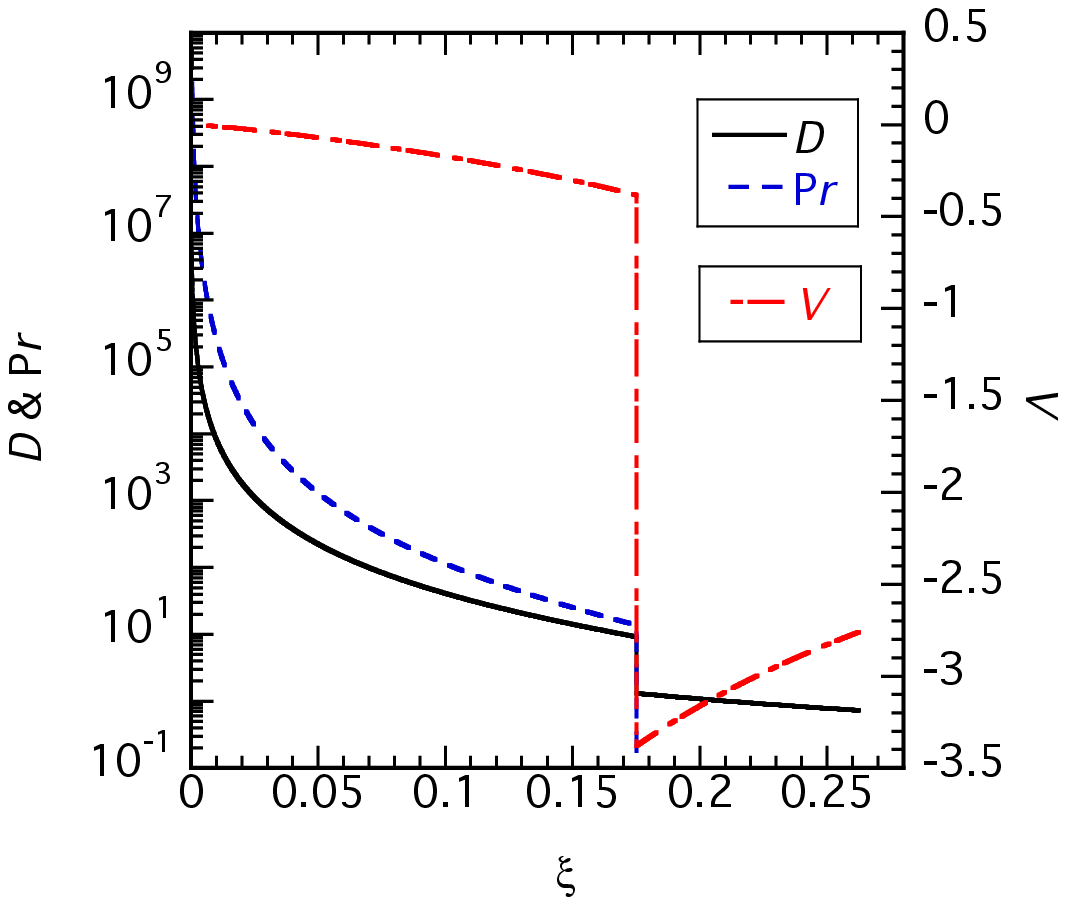} \includegraphics[width=0.45\textwidth]{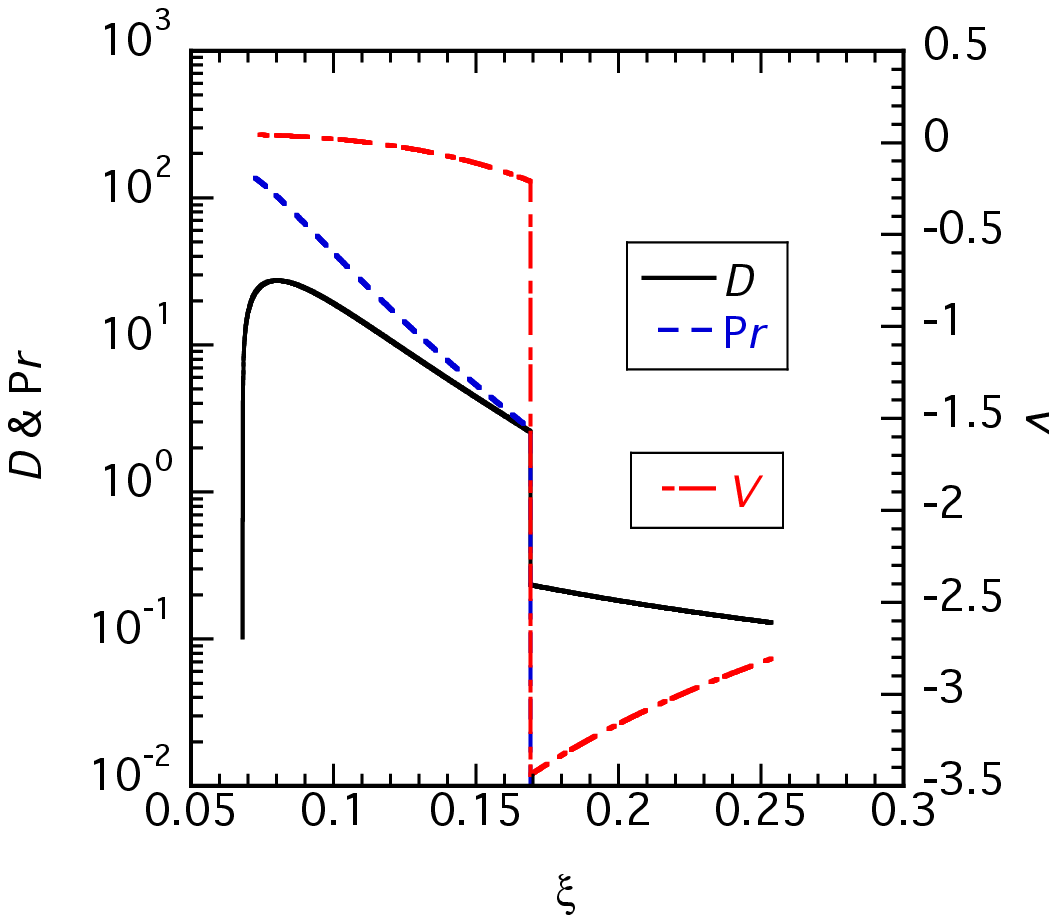}
\caption{Distributions of the density $D$, the pressure $\Pr$, and the velocity $V$ as functions of $\xi$ for solutions with the maximum accretion rates at the shock front. The left panel shows the solution with $\gamma=4/3$ and the right panel $\gamma=6/5$. $l=0$ for both solutions.}
\label{fig:distlt43}
\end{center}
\end{figure}

{Though it might be possible to extend this shock condition to that with an arbitrary strength as was done  in the context of failed supernovae by \citet{2018ApJ...863..158C}, we take the strong shock limit to simplify the procedure to obtain solutions.}

\section{Results}\label{sec:results}
We can find solutions of equations (\ref{eq:ssflc})-(\ref{eq:ssfle}) satisfying the boundary conditions at the shock front and the contact surface for $l>-1$ and $\gamma>1$. 
No solution exists in the other range of these parameters; 
the total deposited energy becomes infinite for $l\leq-1$ and the density becomes negative or infinite at the shock front for $\gamma\leq1$ (see eq. \ref{eq:shockd}). 
We obtain a solution for any positive value of $\xi_{\rm c}$. 
Note that $\xi_{\rm s}$ monotonically increases with $\xi_{\rm c}$ and the solutions in the limit of $\xi_{\rm c}\rightarrow 0$ give the minimum $\xi_{\rm s}$ (denoted by $\xi_{\rm s,\,0}$). 
It follows that there are solutions with the maximum possible accretion rates at the shock front when $\gamma\leq 4/3$ for each $l$. Since the central source is supposed to deposit energy at the rate of $\dot{Q}$, it is convenient to normalize the accretion rate $\dot{M}$ at the shock front with $\dot{Q}$ as 
\begin{equation}\label{eq:rateE}
\frac{GM_{\rm c}\dot{M}}{r_{\rm s}\dot{Q}}=4\pi D_{\rm fb}\sqrt{\xi_{\rm s}},
\end{equation}
where $r_{\rm s}$ denotes the radius of the shock front and $\dot{M}=4\pi r_{\rm s}^2\rho(\xi_{\rm s},\,t)\sqrt{2GM_{\rm c}/r_{\rm s}}$ is the mass accretion rate at the shock front.
Another normalization is possible by taking account of the dimension as
\begin{equation}\label{eq:rate}
\left(\frac{GM_{\rm c}}{t}\right)^{2/3}\frac{\dot{M}}{\dot{Q}}=4\pi D_{\rm fb}\sqrt{\xi_{\rm s}^3}.
\end{equation}
We show these dimensionless accretion rates, $\xi_{\rm s}$, and $D_{\rm fb}$  as functions of $\xi_{\rm c}$ for solutions with $l=0$ and $\gamma=6/5,\, 4/3,\,7/5$, and $5/3$ in Figure \ref{fig:accretion}. Solutions with $\gamma< 4/3$ have maxima of the dimensionless accretion rates given by equation (\ref{eq:rateE}) or (\ref{eq:rate}) at a finite $\xi_{\rm c}$. For $\gamma=4/3$, solutions in the limit of $\xi_{\rm c}\rightarrow 0$ give the maximum accretion rates. Solutions with $\gamma>4/3$ can sustain any accretion rates. In other words, the accretion rate monotonically increases to infinity as $\xi_{\rm c}\rightarrow 0$ in these solutions. At the same time, the value of $D_{\rm fb}$ becomes infinite in this limit.
These characteristic values are listed in Table 1 for some $\gamma$ and $l$.

\begin{table}[hb]
\tbl{Critical dimensionless quantities in solutions with some $l$ and $\gamma$. For $\gamma=5/3$ and $7/5$, values of $\xi_{\rm s,\,0}$ of the solutions in the limit of $\xi_{\rm c}\rightarrow 0$ are listed. For $\gamma=4/3$, in addition to $\xi_{\rm s,\,0}$, the corresponding eigenvalue of $D_{\rm fb}$ (eq. (\ref{eq:shockd})) as well as the corresponding maximum values of the two accretion rates (eqs. (\ref{eq:rateE}) and (\ref{eq:rate})) are listed.  For $\gamma=6/5$, values of $\xi_{\rm s}$ ($\xi_{\rm s,\,1}$ and $\xi_{\rm s,\,2}$) that give the maximum values of the two accretion rates and the corresponding values of $D_{\rm fb}$. }{%
\begin{tabular}{r|l|l|llll|llllll}\hline
&\multicolumn{12}{c}{$\gamma$}\\ \hline
&\multicolumn{1}{c|}{$5/3$} &\multicolumn{1}{c|}{$7/5$}&\multicolumn{4}{c|}{$4/3$}&\multicolumn{6}{c}{$6/5$} \\ \hline
  $l$ & $\xi_{\rm s,\,0}$  &  $\xi_{\rm s,\,0}$  &$\xi_{\rm s,\,0}$  &$ D_{\rm fb,\,0}$ &  $4\pi\left(D_{\rm fb}\sqrt{\xi_{\rm s}}\right)_0$ &  $4\pi\left(D_{\rm fb}\sqrt{\xi_{\rm s}^3}\right)_0$ & $\xi_{\rm s,\,1}$  &$ D_{\rm fb,\,1}$ &  $4\pi\left(D_{\rm fb}\sqrt{\xi_{\rm s}}\right)_1$ & $\xi_{\rm s,\,2}$  &$ D_{\rm fb,\,2}$ &  $4\pi\left(D_{\rm fb}\sqrt{\xi_{\rm s}^3}\right)_2$  \\ \hline \hline
 $-0.99$ & 3.0 &  1.2& 0.75&5.1&55&41 & 0.3007 &4.180&28.81 & 0.6733 & 1.929&13.39\\
$-0.9$ & 1.7 &  0.80& 0.40&2.3&18&7.2 & 0.2839 & 0.5541& 3.710 & 0.6001 & 0.2714& 1.585\\
$-2/3$ & 1.2&0.57&0.28&1.5&10&2.8 & 0.2309 & 0.2908& 1.756& 0.5168 & 0.1358& 0.6341\\
$-0.5$ & 1.1 &0.49&0.25&1.2&7.8&2.0&0.2129&0.2550& 1.479&0.4766&0.1176&0.4862\\
 0 & 0.80  &0.36& 0.18&1.3&6.9&1.2& 0.1691& 0.2332&1.205 &0.4010 & 0.1003&0.3201\\
 1& 0.57& 0.26&0.13  &1.3&5.8 &0.73 & 0.1191 &0.2471&1.072 &0.3170 & 0.09583&0.2149 \\
2 &0.46& 0.20&0.10 &1.4&6.1 & 0.24 & 0.1000& 0.2588&1.029 & 0.2661 &0.09859& 0.1701\\
\hline
\end{tabular}}
\end{table}

Table 1 shows results of solutions with various $l$. The energy deposition rate usually decreases with time if a single physical process is involved and thus described with negative $l$. Such a deposition rate might mimic the declining phase of a sudden energy release due to a glitch in the crust  followed by the transport of energy toward the surface \citep{1989ApJ...336..360E,1991ApJ...381L..47V}. A positive $l$ may be realized when the central object starts  to deposit energy  from a slight excess of the heating rate compared with the cooling rate. Here we present solutions for the energy depositions with linear and quadratic evolutions for simple examples of positive $l$. On the other hand, the solutions with $l=0$ are of particular importance because the magnetic dipole radiation from a pulsar deposits energy at a nearly constant rate in the early phase.   

As mentioned in section \ref{sec:cs}, some solutions have distinct distributions of the density near the contact surface. For example, Figure \ref{fig:dist53} shows solutions with $\gamma=5/3$ and three different values of $l$. The solution with $l=-2/3$ has a finite density at the contact surface, while the density becomes 0 at the contact surface in the solution with $l=0$ and diverges in the solution with $l=-0.9$. The solution with $l=0$ may be subject to the Rayleigh-Taylor instability because the density gradient near the contact surface has a sign opposite to that of the pressure gradient  (see the left panel of Fig. \ref{fig:dist53}). This instability tends to bring dense matter toward the contact surface. Thus the central activity may fail to repel the fallback matter even if $\gamma>4/3$.

Solutions with the maximum accretion rates are shown in Figure \ref{fig:distlt43}. These solutions have negative velocities in some part of Region II because of their small $\xi_{\rm s}$'s.  From equation (\ref{eq:shockv}), the velocity immediately behind the shock front becomes negative when $\xi_{\rm s}<\left[3(\gamma-1)\right]^{2/3}/2$. If $\gamma=4/3$ ($\gamma=6/5$),  this criterion becomes $\xi_{\rm s}<1/2$ ($\xi_{\rm s}<(5/3)^{2/3}/2\sim0.356$).  On the other hand, the solutions with $\gamma=5/3$ presented in Figure \ref{fig:dist53} happen to have large $\xi_{\rm s}$'s that do not satisfy this criterion ($\xi_{\rm s}<1/2^{1/3}$ for $\gamma=5/3$), thus the shocked matter flows outward  while solutions with $\gamma>4/3$ and negative $l$ have $\xi_{\rm s,\,0}$ greater than the upper limit of the criterion as seen in Table 1. Shocked matter in a solution with a positive $l$ tends to flow inward due to the weak central engine in the early phase. Because the contact surface always moves outward in all the solutions (see eq. (\ref{eq:contact})), the pressure increased by the central energy deposition repels the fallback matter somewhere in Region II even for these solutions with great accretion rates.  In solutions with small accretion rates, the energy deposition can repel the fallback at the shock front.

{The kinetic energy $E_{\rm k}(t)$ of the repelled ejecta evolves with time $t$ following the equation
\begin{equation}
E_{\rm k}(t) =2\pi L_lt^{l+1}\int_{\xi_{\rm c}}^{\xi_{\rm s}}D(\xi)V^2(\xi)d\xi.
\end{equation}
Since the value of the dimensionless integral is of the order of unity here, the factor in front of the integral gives a rough amount of the kinetic energy.}

\section{Application to neutron star formation with fallback accretion}
Although the self-similar solutions constructed in the previous sections are valid under limited conditions, they can be practically applicable to some astrophysical phenomena. 
Here we mainly focus on supernova fallback pushed back by the energy deposition from a newborn neutron star, in particular, a rotation-powered one.
We discuss the conditions for repelling the fallback accretion by the spin-down luminosity and their implications on properties of associated supernovae and fates of the neutron stars.  

After a successful supernova shock propagates through the progenitor star, a bulk of the stellar material is ejected while a minor fraction falls back. 
The accretion rate and its temporal evolution can be determined by the strength of the supernova shock and the inner structure of the progenitor. 
For example, \citet{2016ApJ...821...69E} estimated the fallback rate based on one-dimensional numerical simulations of neutrino-driven explosions. 
The fallback typically starts when the neutrino luminosity significantly decreases, i.e., $t_{\rm fb} \gtrsim 10$ s after the core bounce, 
and the accretion rate subsequently decreases as $\propto (t/t_{\rm fb})^{-5/3}$ for $t \gtrsim t_{\rm fb}$. 
The total mass $M_{\rm fb}$ of the fallback matter ranges from $ \sim 10^{-4}\,M_\odot$ to $\sim 10^{-2}\,M_\odot$ . More matter falls back in a more massive star.
Correspondingly, the peak mass accretion rate is estimated to be $\dot M_{\rm fb} \lesssim 10^{-(3\mbox{-}5)}\,M_\odot\,\rm s^{-1}$. 

In general, the maximum fallback accretion rate that can be repelled by the central energy source is described with
$M_{\rm c}$, $r_{\rm s}$, $(4 \pi D_{\rm fb} \sqrt{\xi_{\rm s}})$, and $\dot Q$ from equation (\ref{eq:rateE}). 
Since we consider a fallback accretion onto a neutron star, we take $M_{\rm c} = M_* \sim 1.4 \ M_\odot$ and set the radius of the neutron star as $R_* \sim 12 \ \rm km$. 
When the fallback accretion starts, the position of the shock will be $r_{\rm s} \approx \xi_{\rm s} (GM_*t_{\rm fb}{}^2)^{1/3}$, or 
\begin{equation}\label{eq:r_s_crit}
r_{\rm s} \sim 4.2 \times 10^{9}\,{\rm cm}~\xi_{\rm s}\,\left(\frac{t_{\rm fb}}{20\,\rm s} \right)^{2/3}.
\end{equation}
In the critical situation corresponding to the maximum mass accretion rate, the radius of the inner contact surface should be set as the neutron star radius, i.e., $\xi_{\rm c, crit} \approx R_*/(GM_*t_{\rm fb}{}^2)^{1/3}$, or 
\begin{equation}\label{eq:xi_c_crit}
\xi_{\rm c, crit} \sim 2.8\times 10^{-4}\,\left(\frac{t_{\rm fb}}{20\,\rm s} \right)^{-2/3}.
\end{equation}
The temperature in the shocked region is typically high and the radiation pressure dominates so that the adiabatic index will be slightly larger than $4/3$. 
Combining this fact with equation (\ref{eq:xi_c_crit}), we estimate the critical values of $\xi_{\rm s}$ and $(4 \pi D_{\rm fb} \sqrt{\xi_{\rm s}})$ from Figure \ref{fig:accretion} as {
\begin{equation}
\xi_{\rm s, crit} \sim 0.2\ {\rm and}\ (4 \pi D_{\rm fb} \sqrt{\xi_{\rm s}})_{\rm crit} \sim 5.3	. 
\end{equation}
}
Hereafter we will assume $l = 0$ which is typically valid for $t \sim t_{\rm fb} \gtrsim 10\,\rm s$ since the timescale $t_{\rm sd}$ of the spin-down is estimated to be much longer than that of the fallback from the following formula:
\begin{equation}\label{eq:sdtime}
t_{\rm sd}=-\left(\frac{\Omega}{\dot\Omega}\right)\sim\frac{6Ic^3}{B_*^2R_*^6\Omega^2}\sim2\times10^9~{\rm s}\left(\frac{B_*}{10^{13}~{\rm G}}\right)^{-2}\left(\frac{P}{10~{\rm ms}}\right)^2,
\end{equation}
where $I$ denotes the moment of inertia of the neutron star, $\Omega(=2\pi/P)$ the angular frequency ($P$ the spin period), $B_*$ the  magnetic field at the magnetic pole on the surface, and $c$ denotes the speed of light.
Finally, the maximum energy deposition rate from a rotating neutron star can be described as $\dot Q_{\rm crit} \approx (\mu^2 \Omega^4/c^3) \times (R_{\rm lc}/R_*)^2$, or 
\begin{equation}\label{eq:Q_crit}
\dot Q_{\rm crit} \sim 2.7 \times 10^{45}\,{\rm erg\,s^{-1}}\,\left(\frac{B_*}{10^{13}\,\rm G} \right)^2\left(\frac{P}{10\,\rm ms} \right)^{-2},
\end{equation}
where $\mu = B_*{}R_*{}^3$ is the magnetic dipole moment and $R_{\rm lc} = c/\Omega$ is the light cylinder radius.  
We note that the above energy deposition rate is different from the classical dipole spin-down rate, $\dot Q_{\rm dipole} \approx  (\mu^2 \Omega^4/c^3)$.
The additional factor $(R_{\rm lc}/R_*)^2$ represents the enhancement of spin-down luminosity~\citep{2016ApJ...822...33P}; 
the magnetic fields are maximally open, like a split monopole, due to the accretion.  
From equations (\ref{eq:r_s_crit}-\ref{eq:Q_crit}), we obtain the critical accretion rate as {
\begin{equation}\label{eq:Mdot_crit_ns}
\dot M_{\rm crit, repul} \sim 3 \times 10^{-5}\,M_\odot\,{\rm s^{-1}}\,\frac{\xi_{\rm s, crit}}{0.2}\frac{(4 \pi D_{\rm fb} \sqrt{\xi_{\rm s}})_{\rm crit}}{5.3} \left(\frac{B_*}{10^{13}\,\rm G} \right)^2\left(\frac{P}{10\,\rm ms} \right)^{-2}\left(\frac{t_{\rm fb}}{20\,\rm s} \right)^{2/3}.
\end{equation}}
If $\dot M_{\rm fb} \lesssim \dot M_{\rm crit, repul}$, a bulk of the fallback matter will be directly repelled by the spin-down power. 
Otherwise, it is accreted on the neutron-star surface. 

If $\dot M_{\rm fb} \gtrsim \dot M_{\rm crit, repul}$, the fallback matter reaches the neutron star surface. 
If the accretion rate is so large, then the surface magnetic field can be buried and the spin-down power is significantly reduced~(e.g., \cite{2010RMxAA..46..309B,2016MNRAS.456.3813T}). 
A necessary condition for burying the magnetic field is set by the balance between the magnetic pressure and the ram pressure at the surface; 
\begin{equation}
\frac{B_*{ }^2}{8\pi}\lesssim\rho v^2\sim\frac{\dot M}{4\pi R_*^2}\sqrt{\frac{GM_{\rm c}}{R_*}},
\end{equation}
where $\rho$ and $v$ are the density and velocity of the accreted matter near the surface.  The second equality assumes the steady accretion of freely falling matter. Numerically, this yields the critical accretion rate of
\begin{equation}
\dot M_{\rm crit, bury} \sim 3\times10^{-6}\,M_\odot\,{\rm s^{-1}}\,\left(\frac{B_*}{10^{13}\,{\rm G}}\right)^2. 
\end{equation}
Whether the magnetic fields are actually buried into the crust may be addressed by comparing the position of the magnetopause and the crust radius.  
In this way, \citet{2016MNRAS.456.3813T}\footnote{There is a typo in their eq. (25).} estimated the threshold value as 
\begin{equation}
\dot M_{\rm crit, bury} \sim 10^{-5}\,M_\odot\,{\rm s^{-1}}\,\left(\frac{B_*}{10^{13}\,{\rm G}}\right)^{3/2}. 
\end{equation}

\begin{figure}
  \includegraphics[width=0.32\textwidth]{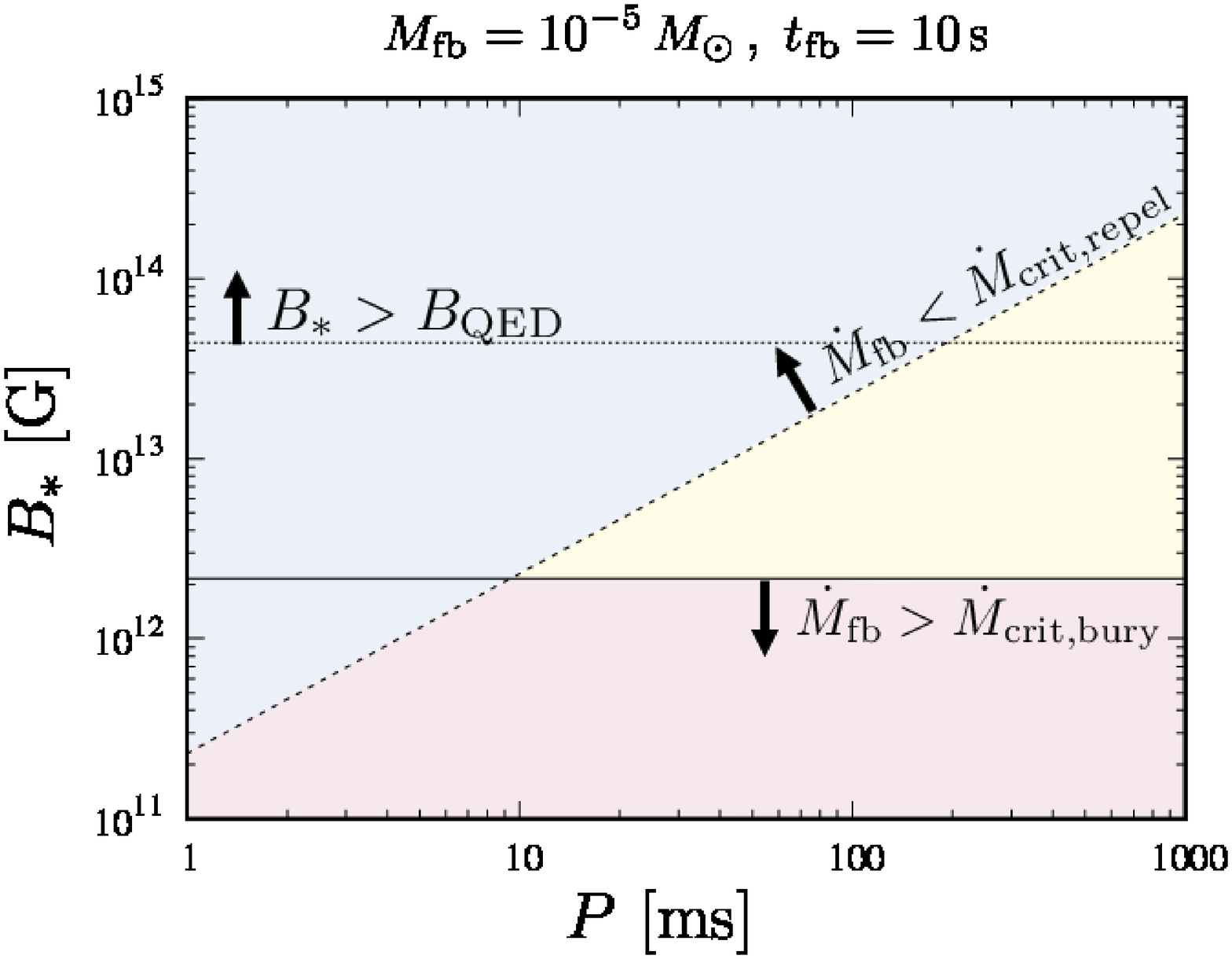} 
  \includegraphics[width=0.32\textwidth]{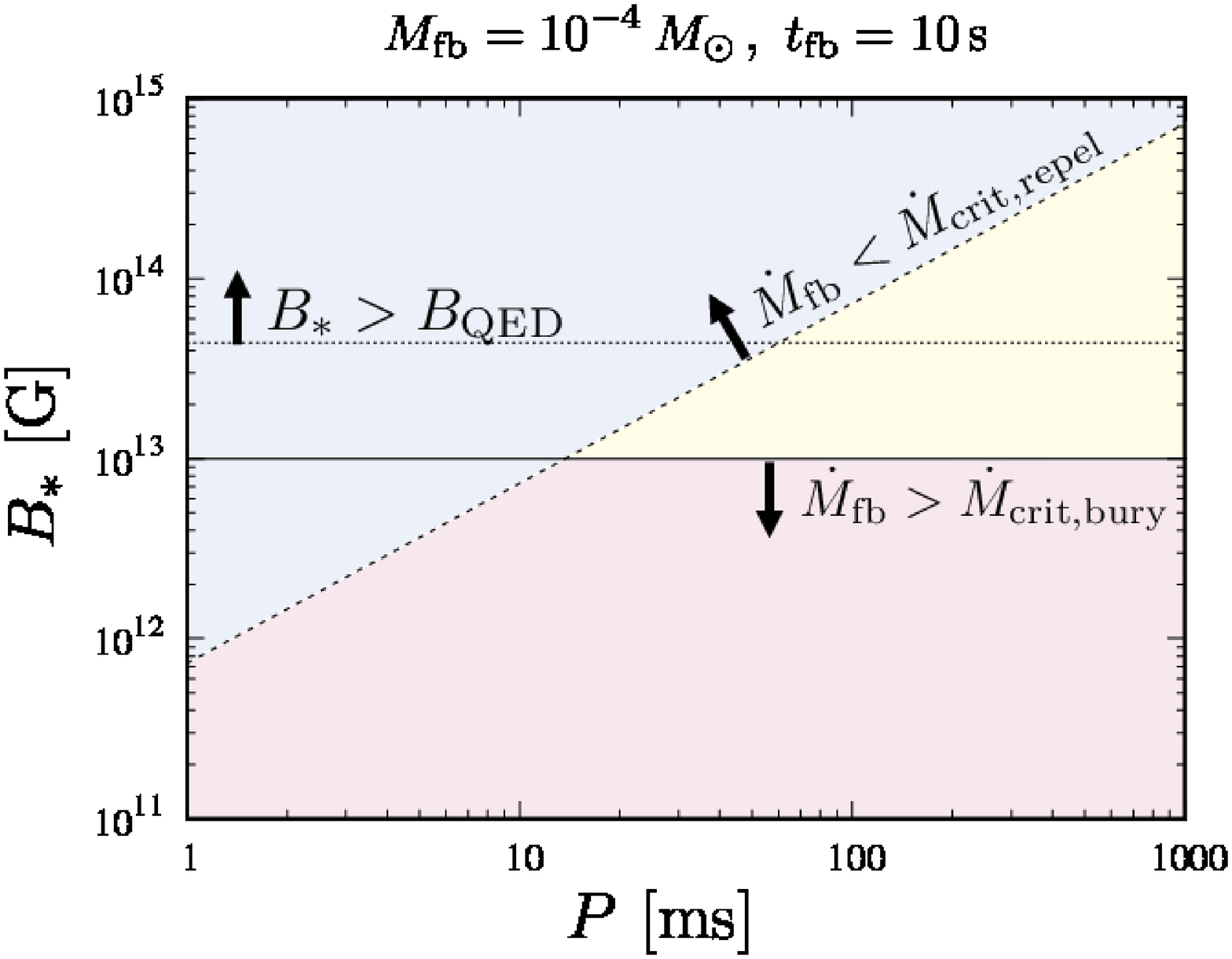} 
  \includegraphics[width=0.32\textwidth]{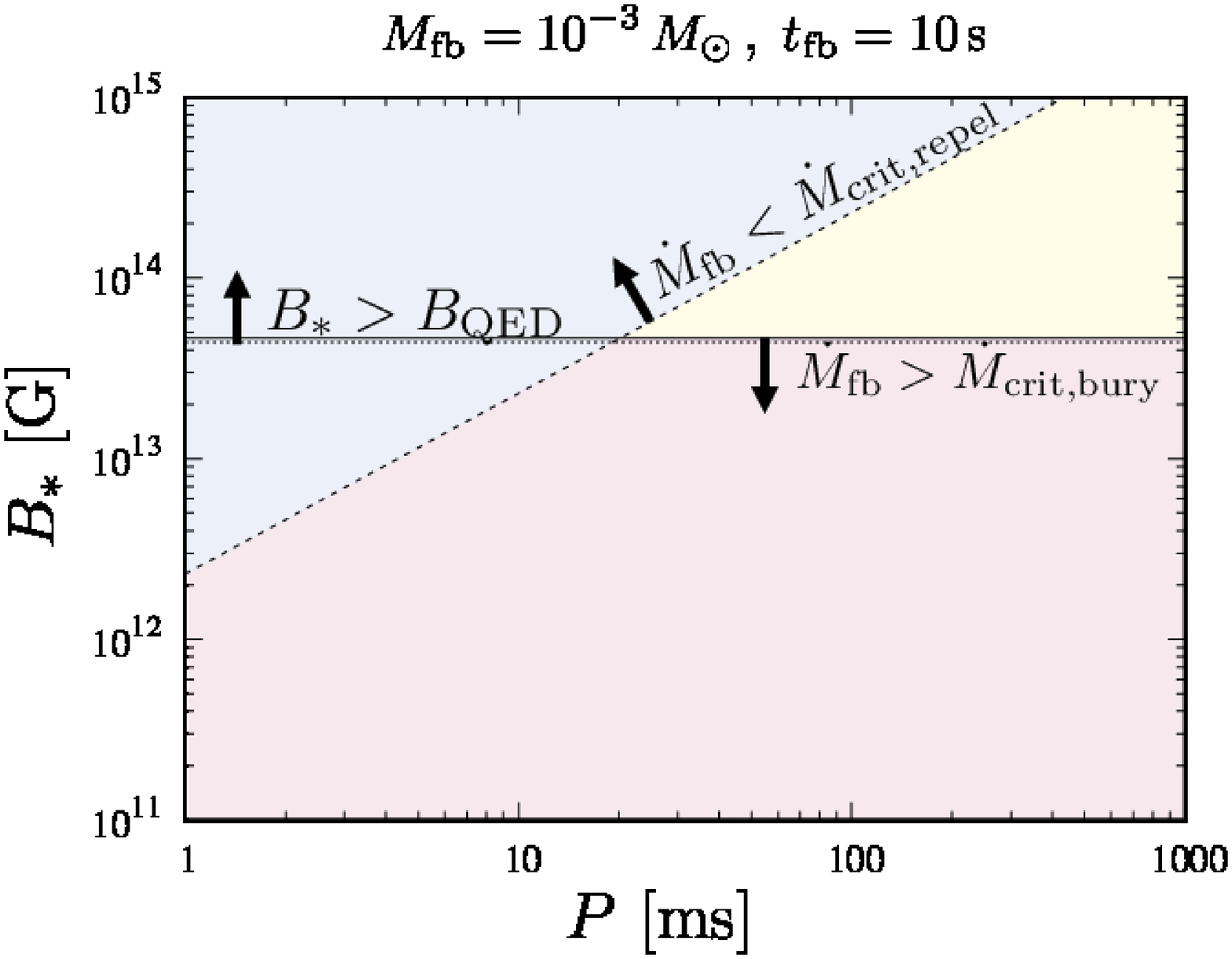} 
\caption{
Consequences of fallback accretion onto a neutron star and their dependencies on the magnetic field and rotation.
The left, center, and right panels show the cases of  $\dot M_{\rm fb} = 10^{-6}\,M_\odot\,\rm s^{-1}$, $10^{-5}\,M_\odot\,\rm s^{-1}$, and $10^{-4}\,M_\odot\,\rm s^{-1}$, respectively. 
}
\label{fig:cco_psr}
\end{figure}

In the case of $\dot M_{\rm crit, repul} \lesssim \dot M_{\rm fb} \lesssim \dot M_{\rm crit, bury}$, the situation will be more complicated. 
At first, the fallback matter can be accreted on the neutron star; the magnetic fields and spin-down energy are confined in a near surface region.
As the fallback rate decreases with time, large-scale fields emerge and the spin-down power pushes back the fallback matter.
In Figure \ref{fig:cco_psr}, we show how the consequences of a fallback accretion depend on $B_*$ and $P$ for   
$\dot M_{\rm fb} = 10^{-6}\,M_\odot\,\rm s^{-1}$(left), $10^{-5}\,M_\odot\,\rm s^{-1}$(center), and $10^{-4}\,M_\odot\,\rm s^{-1}$(right). 

Let us now discuss possible connection between the diversities of neutron-star formation with fallback accretion and the observed young neutron stars. 
From Figure \ref{fig:cco_psr}, the condition $\dot M_{\rm fb} < \dot M_{\rm crit, repul}$ is always satisfied 
for fast-spinning strongly-magnetized neutron stars with $B_* \gtrsim 10^{13}\,\rm G$ and $P \lesssim$ a few ms. 
Such cases have been proposed as a plausible central engine of extragalactic transients like gamma-ray bursts, superluminous supernovae, and fast radio bursts~(see e.g., \cite{2015MNRAS.454.3311M,2016ApJ...818...94K,2017ApJ...839L...3K,2018MNRAS.475.2659M} and references therein)~\footnote{Also see \citet{2018ApJ...857...95M} for the impact of fallback accretion on the time evolution of spin in the early stage.}.  
A similar range of $B_*$ and $P$ has been also considered in the context of Galactic magnetar formation; 
the magnetic field amplification can be attributed to the proto-neutron-star convection coupled with a differential rotation~\citep{1992ApJ...392L...9D,1993ApJ...408..194T} 
or the magneto-rotational instability~(e.g., \cite{2015Natur.528..376M}). 
Note, however, that so far there is no observational support to the dynamo scenario~(e.g., \cite{2006MNRAS.370L..14V}).

Neutron stars with relatively weak magnetic fields ($B_* \lesssim 10^{13}\,{\rm G}\,(\dot M_{\rm fb}/10^{-5}\,M_\odot\,\rm s^{-1})^{2/3}$) and slow spins ($P \gtrsim$ a few 10 ms) satisfy
the condition $\dot M_{\rm fb} > \dot M_{\rm crit, bury}$ . 
Both the magnetic field and spin-down power are strongly suppressed by the accreted matter.  
As long as the situation continues, the thermal energy stored in the neutron star will be the main source of the emission. 
Such neutron stars will be observed as CCOs as proposed by \citet{2016MNRAS.456.3813T}. 

The boundary between the former and latter cases is set by  equation (\ref{eq:Mdot_crit_ns}), 
i.e., $(B_*/10^{13}\,{\rm G}) \times (P/10\,{\rm ms})^{-1} \sim (\dot M_{\rm fb}/10^{-4}\,M_\odot\,\rm s^{-1})^{1/2}$, 
which might correspond to the boundary between rotation-powered pulsars and CCOs. 
For example, the initial magnetic field and spin period of the Crab pulsar have been estimated to be $B_* \sim 10^{13}\,\rm G$ and $P =$ a few 10 ms, respectively, 
and consistent with the above point if $\dot M_{\rm fb} \lesssim 10^{-5}\,M_\odot\,\rm s^{-1}$. 
Such a relatively small fallback accretion rate might be also consistent with a relatively low-mass progenitor inferred for the Crab pulsar. 

In the intermediate cases where 
$(\dot M_{\rm fb}/10^{-5}\,M_\odot\,\rm s^{-1})^{2/3} \lesssim (B_*/10^{13}\,{\rm G}) \lesssim (P/10\,{\rm ms})\times(\dot M_{\rm fb}/10^{-4}\,M_\odot\,\rm s^{-1})^{1/2}$, 
the magnetic field eventually emerges and the energy source of the emission can be either spin-down power or decay of the magnetic-field. 
Interestingly, the parameter range of $B_* \sim B_{\rm QED} = 4.4 \times 10^{13}\,\rm G$ and $P \gtrsim$ a few 10 ms, especially for $\dot M_{\rm fb} \gtrsim 10^{-5}\,M_\odot\,\rm s^{-1}$
is somewhat consistent with the so-called fossil scenario of high- and low-field magnetar formation~\citep{2006MNRAS.367.1323F}. 
Such a relatively large fallback accretion rate might be expected in relatively high-mass progenitors inferred for magnetars 
(e.g., \cite{2005ApJ...622L..49F,2005ApJ...620L..95G,2006ApJ...636L..41M,2014ApJS..212....6O}). 

As above, we show that the observed diversity of Galactic neutron stars could be connected to $(B_*, P, \dot M_{\rm fb})$ at their birth. 
We speculate that ordinary pulsars, CCOs, and magnetars originate from the blue, red, and yellow shaded regions in Figure \ref{fig:cco_psr}, respectively. 
The fact that the three classes have a comparable population (e.g., \cite{2008MNRAS.391.2009K}) can be naturally explained since the typical parameters at their birth, $B_* \sim 10^{13}$ G and $P \sim$ a few 10 ms,
roughly coincide with the intersection of the boundaries of three regions. 
We should note, however, that the self-similar solutions in Sec. \ref{sec:results} and calculations by \citet{2016MNRAS.456.3813T} are one dimensional. 
Multi-dimensional effects in fallback accretion have to be taken into account consistently to make the above criterions more quantitatively accurate. {For instance, as mentioned in section \ref{sec:results},  solutions with $l=0$ might be subject to the Rayleigh-Taylor instability. Nevertheless, multi-dimensional effects are not expected to change the above criterions because the energy of the shocked flow in Region II is always positive. Of course, one needs to perform multi-dimensional calculations to know detailed outcomes of this instability.}
We should also note that, in order to connect their states at birth to observational properties at $t_{\rm age} \gtrsim $ a few 100 yrs, the long-term evolution of spins and magnetic fields become important. 
We will investigate these topics elsewhere.

\section{Summary and Discussion}\label{sec:discussion}
We have presented series of self-similar solutions for fallback matter being shocked and repelled by the energy deposition from a central object. 
The behavior of the solutions changes depending on the adiabatic index $\gamma$ in the shocked region. 
For $\gamma > 4/3$, we can find a self-similar solution for an arbitrarily high fallback accretion rate by taking the radius of the contact surface correspondingly small.
On the other hand, for $\gamma \leq 4/3$, there are upper-bounds to the accretion rate at the shock front $r = r_{\rm s}$:
\begin{equation}\label{eq:Mdot_crit}
\dot M_{\rm crit} = \frac{r_{\rm s} \dot Q}{G M_{\rm c}} \times (4 \pi D_{\rm fb} \sqrt{\xi_{\rm s}})_{\rm crit}. 
\end{equation}
The critical values of $4 \pi D_{\rm fb} \sqrt{\xi_{\rm s}}$, ranging from $\sim 1-100$, are shown in Table 1.
Note that $4 \pi D_{\rm fb} \sqrt{\xi_{\rm s}} = 1$ corresponds to the cases 
where the accretion luminosity $G \dot M M_{\rm c}/r_{\rm s}$ is equal to the energy deposition rate from the central object $\dot Q$. 
For $\dot M > \dot M_{\rm crit}$, the fallback matter will plunge into the central object.

We have applied the self-similar solution to neutron star formation with fallback accretion when the neutron star deposits energy due to its spin-down power. The solution suggests that the critical mass accretion rate above which the fallback matter can be accreted on the surface is given as a function of  the initial spin rate and surface magnetic filed.  It is shown that this condition together with the criterion that the accreted matter can bury the magnetic field and suppress the spin-down power may determine the fate of the newborn neutron star to become a magnetar, a pulsar, or a CCO.

\subsection{Relation to Previous Works}
We here discuss relations between the self-similar solution constructed in this paper and previous studies on the dynamics of fallback accretion.  

The shock propagates outward with the radius proportional to $t^{2/3}$ in our models. A similar behavior of the shock front was obtained from numerical computations for fallback presented in \citet{1972SvA....16..209Z}. Thus our model can capture some features of realistic models that treat the energy source originating from gravitational energy of accreted matter. \citet{1972SvA....16..209Z} discussed the accretion of up to 10$^{-5}$ $M_\odot$ assuming a static power-law density distribution proportional to $r^{-3/2}$ ($r$ denotes the radial coordinate) as the initial condition, which results in a constant accretion rate. They computed hydrodynamics of the fallback matter including radiative diffusion of photons and neutrino emission. \citet{1971ApJ...163..221C} emphasized the role of neutrino emission in driving intense fallback motion.  These investigations focused on the early phase affected by the unknown explosion mechanism of a core collapse supernova. 

\citet{1989ApJ...346..847C} discussed the fallback phenomenon after a reverse shock wave reaches the neutron star.  He argued that the accretion rate of the bound debris decreases with time $t$ following a power law as $t^{-5/3}$ provided that the mass fraction of the debris is uniformly distributed in binding energy. This power law evolution of the accretion rate had been already pointed out from a dimensional analysis \citep{1988Natur.333..644M}. Though \citet{1989ApJ...346..847C} started his computations from the uniform motion of uniform matter to obtain this accretion rate,  he pointed out that this situation can be realized if the density $\rho_{\rm d}(r)$ of the debris initially has a static power-law distribution $\rho_{\rm d}(r)\propto r^{-4}$ \citep{1974Ap&SS..31..251S}.  He also discussed the effects of the pulsar activity on the fallback and concluded that the effects are negligible if only the magnetic pressure is considered. 

On the contrary, newborn spinning neutron stars should deposit energy in the surrounding matter as a result of the activities originating from rotating strong magnetic fields. This interaction has been discussed with self-similar solutions in 1D spherical symmetry \citep{2005ApJ...619..839C, 2017MNRAS.466.2633S} and numerical simulations in 2D axi-symmetry \citep{2016ApJ...832...73C, 2017MNRAS.466.2633S, 2017ApJ...845..139B} or 3D \citep{2017ApJ...845..139B}. All of these studies have dealt with interactions between pulsar winds and expanding supernova ejecta without fallback matter. In reality, the innermost part of once ejected matter falls back and the  energy deposited by the pulsar wind should affect the motion of the fallback matter, which is important for determining the nature of young neutron stars as discussed in section 4. Our solutions cover some of these aspects though we simplified the situation by introducing the power law energy deposition rate and assuming a specific motion of the fallback matter above the shock front, which is different from the assumptions in the above previous work. Thus our solutions have a different temporal evolution of the mass accretion rate.

\begin{ack}
This work is partially supported by JSPS KAKENHI Grant Number JP17K14248, JP18H04573, 16H06341, 16K05287, 15H02082, MEXT, Japan.
\end{ack}

\end{document}